# Towards a frequency independent incremental ab initio scheme for the self energy

M. Albrecht*

*Theoretical Chemistry FB08, University of Siegen, 57068 Siegen/Germany*

## Abstract

The frequency dependence of the self energy of a general many–body problem is identified as a main obstacle in correlation calculations based on local approaches. A frequency independent formulation is proposed instead and proven to yield exactly the same numerical results as the original common scheme. Our approach is embedded in a general local–orbital based *ab initio* frame to obtain the Green's function for large heterogenous systems. First a Green's function formalism is introduced. Then the self energy is constructed from an incremental scheme. Subsequently we apply the proposed frequency independent formulation. The theory is applied to para–ditholbenzene as a realistic system and the numerical accuracy of the correlation contributions obtained from our frequency independent access are carefully tested against the exact frequency dependent results. Perfect agreement is reached and a speed–up of a factor 50 is established for the incremental scheme.

* Corresponding author: e-mail: m.albrecht@uni-siegen.de, Phone: +49 271 740 4024, Fax: +00 271 740 2805

## I. INTRODUCTION

The field of many–body theory has been developing at breathtaking speed in recent years. A significant amount of effort is directed towards affordable descriptions of electronic correlation in infinite periodic systems or large heterogeneous systems both in the ground state and in excited states. Well established schemes like the density functional theory (DFT)[1] have seen more and more refinements, for example by means of the optimized effective potential method (OEP)[2], time dependent DFT[3], screening implementations[4], the Wigner theory[5] or the reduced density matrix functional theory (RDMFT)[6]. Other correlation schemes have been derived, examples are the dynamical mean field theory (DMFT)[7] or the GW approximation[8].

Parallel to the density functional approaches wave function based quantum chemical methods have also been trimmed to be applicable to large systems and finally even to polymers and crystals. As for ground state properties the ansatz of a local incremental scheme in combination with the coupled cluster (CC) method turned out to be valiant in applications to large molecules or polymers and solids[9–14]. Significant advances for the case of excited states have been reported for quantum monte carlo methods (QMC)[15–18], algebraic diagrammatic construction (ADC)[19], effective Hamiltonians[20,21] or the Green's function approach[22–27].

These wave function based methods are straightforwardly applicable to both ground state and excited state calculations alike and are amenable to systematic improvement on the numerical accuracy by their very construction.

The general bottle–neck of steep increase of numerical effort with system size, however, affects all wave function based methods alike. It is precisely this obstacle which was overcome in earlier applications by a formulation of electron correlations in local orbitals and a hierarchy of correlation contributions called the incremental scheme[9,13,28–30].

Recently the author also demonstrated, that a full Green's function approach with a frequency dependent self energy can in principle also be combined with the incremental scheme. Band structure calculations were performed for LiH and LiF[24–26] and a recent application to a molecule inside a molecular junction also underlines the usefulness of such an approach[27]. The key enabling such calculations was an approach based on local orbitals and a real space formulation of the self energy. However, at the same time it became manifest that the explicit frequency dependence of the self energy prevents a full exploitation of the potential of the incremental scheme in local orbitals, which was tremendously successful in applications to frequency independent quantities.

In the present work we device a path around this problem. The application of a mathematical identity allows to calculate a frequency independent intermediate quantity which can be obtained by the incremental scheme in very much the same way as the effective Hamiltonian or the ground state correlation energy was obtained before. The full frequency dependent self energy is recovered *a posteriori* without any additional numerical effort. In this way we can finally obtain the full frequency dependent self energy and hence the Green's function with a maximum of numerical efficiency. This new approach overcomes the main obstacle in applications of the incremental scheme to the Green's function and we are thus led to believe that this constitutes a very important step ahead.

In the following section we describe the theory with the focus on the transformation from a frequency dependent to an independent scheme. In Sec. III several numerical tests are presented for a realistic system and Sec. IV contains our conclusions.

## II. THEORY

In earlier works we have designed a formulation of the Green's function correlation method so as to use local HF orbitals as a starting point and then assess the correlation contributions in a full *ab initio* manner. This correlation method has subsequently been combined with the so called incremental scheme which proved to boost efficiency tremendously[24–27].

However, in the case of the Green's function theory correlations are described with a frequency dependent quantity, the self energy. This is different from all earlier applications of the incremental scheme to ground state properties[9–11,13,14] or excited states in solids[20,21,29,30]. So far this has prevented the method from exploiting the full computational power of the incremental scheme in local orbitals.

We first briefly repeat the Green's function correlation method and then give a description of the incremental scheme in Sec. II A, II B.

In the main parts Sec. II C, II D a mathematical equality is presented to overcome the numerical obstacle of the frequency dependence. The quantum chemical expressions are adapted towards this formula and several numerical checks are perfomred in Sec. III.

### A. The Green's function

The starting point of our approach are localized Hartree–Fock (HF) orbitals for the system at hand. This could be a medium sized or large molecule or even a polymer or a three–dimensional crystal. The following formulas are kept completely general, the main issue being that the starting orbitals be local ones. In terms of such orbitals a model space P and excitation space Q are distinguished for the example of virtual states (the case of occupied states being completely analogous) as follows: The model space P describing the HF level comprises of the $(N + 1)$–particle HF determinants $|\eta\rangle$, while the correlation space Q contains single and double excitations $|\alpha\rangle$ on top of $|\eta\rangle$:

$$|\eta\rangle = c_n^\dagger |\Phi_{\mathrm{HF}}\rangle, \qquad |\alpha\rangle = c_r^\dagger c_a |\eta\rangle, \quad c_r^\dagger c_s^\dagger c_a c_b |\eta\rangle \tag{1}$$

$$P = \sum_\eta |\eta\rangle\langle\eta|, \qquad Q = \sum_\alpha |\alpha\rangle\langle\alpha|. \tag{2}$$

We adopt the index convention that $a, b, c, d$ and $r, s, t, u$ represent occupied and virtual local HF orbitals, respectively. (In the case of infinite periodic systems these would be composed indices comprising of a lattice vector, an orbital index and a spin index.) The idea of locality translates into a restriction of the area the orbital can be chosen from to one or more contiguous spatial parts of the system. It is important to note that by enlarging the size of the spatial area thus covered, this approximation can be checked in a systematic way for convergence. A very important implication of this procedure is that the correlation methods employed must be size–consistent. This idea is elaborated upon in Sec. II B.

The Green's function approach has precisely the merit of being intrinsically size–consistent, so that a diagonalization allows to go beyond perturbative results. Pertaining to the above notation the Green's function $G_{\mathrm{nm}}(t) = -i\langle T[c_{\mathrm{n}}(0)c_{\mathrm{m}}^\dagger(t)]\rangle$, where T is the time–ordering operator and the brackets denote the average over the exact ground–state, can be obtained from Dyson's equation as:

$$G_{nm}(\omega) = [\omega - F - \Sigma(\omega)]_{\mathrm{nm}}^{-1}. \tag{3}$$

Here the self energy $\Sigma_{\mathrm{nm}}(\omega)$ which contains the correlation effects, has been introduced. $G_{\mathrm{nm}}^0(\omega)$ is the HF propagator $[G^0(\omega)]_{\mathrm{nm}}^{-1} = \omega - F_{\mathrm{nm}}$. The correlated energies are given by the poles of the Green's function which are numerically iteratively retrieved as the zeros of the denominator in Eq. (3). To construct the self energy the resolvent

$$[\omega - H^{\mathrm{R}} + i\delta]_{\alpha;\alpha'}^{-1} \tag{4}$$

is needed. It can be gained from diagonalization of the Hamiltonian

$$[H^{\text{R}}]_{\alpha,\alpha'} = \langle\alpha|H - E_0|\alpha'\rangle, \tag{5}$$

where the states $|\alpha\rangle, |\alpha'\rangle$ are those of the correlation space Q as in Eq. (1). Here $E_0$ is the HF ground state energy while the brackets indicate the HF average.

Along with a straightforward implementation several perturbative approximations have been derived and analyzed. Theoretical connections to the perturbative effective Hamiltonian[20] were also established as mentioned below.

The self energy is approximated by decomposition into a retarded and an advanced part

$$\Sigma_{\text{nm}}(\omega) = \Sigma_{\text{nm}}^{\text{R}}(\omega) + \Sigma_{\text{nm}}^{\text{A}}(\omega). \tag{6}$$

In what follows the superscript $^{\text{R}}$ will be used throughout to refer to the retarded case, while $^{\text{A}}$ is taken to denote the advanced part. Furthermore, the configuration space will be restricted to single excitations, i. e. three–body–interactions.

In the following only the construction of the retarded self energy part is given, the case of the advanced part being analogous.

The space of 2–particle 1–hole states (2p1h) is spanned by $|r, s, a\rangle = a_r^\dagger a_s^\dagger a_a|\Phi_{\text{HF}}\rangle$. The Hamiltonian is set up in this basis as: $[H^{\text{R}}]_{\text{rsa},r's'a'} = \langle r, s, a|H - E_0|r', s', a'\rangle$ and is subsequently diagonalized.

Diagonalizing the matrix $H^{\text{R}}$ results in the eigenvectors $S^{\text{R}}$ and eigenvalues $\lambda^{\text{R}}$. The retarded part of the self energy is then constructed as

$$\begin{aligned}
\Sigma_{\text{nm}}^{\text{R}}(\omega) &= \sum_{\text{rsa};r's'a'} \Gamma(\text{rs};\text{na})\left[\omega - H^{\text{R}} + i\delta\right]^{-1}_{\text{rsa};r's'a'}\Gamma(\text{r's'};\text{ma'}) \\
&= \sum_{\text{rsa};r's'a'} \Gamma(\text{rs};\text{na}) \sum_q S_{rsa;q}^{\text{R}} \frac{1}{\left(\omega - \lambda_q^{\text{R}} + i\delta\right)} S_{q;r's'a'}^{\text{R}} \Gamma(\text{r's'};\text{ma'}).
\end{aligned} \tag{7}$$

$\Gamma$ is a shorthand for $\Gamma(\text{rs};\text{ta}) = W_{\text{rsta}} - W_{\text{rsat}}$ and $W$ are the standard two–electron integrals. Numerically it is advantageous to first form and store the frequency independent quantity

$$D_{\text{n};q}^{\text{R}} = \sum_{rsa} \Gamma(\text{rs};\text{na}) S_{rsa;q}^{\text{R}} \tag{8}$$

and rewrite expression (7) as

$$\Sigma_{\text{nm}}^{\text{R}}(\omega) = \sum_q D_{\text{n};q}^{\text{R}} \frac{1}{\left(\omega - \lambda_q^{\text{R}} + i\delta\right)} D_{\text{q};m}^{\text{R}}. \tag{9}$$

### B. The incremental scheme

The efficiency of the procedure is derived from the application of an incremental scheme. The task of correlating electrons in a large system is broken down systematically to diagonalizations in smaller subsystems. A sketch of the idea is given in Fig. 2 in Sec. III.

As subsets of the incremental scheme some arbitrary spatial parts of a molecule, representing a suitable partitioning, are chosen for this example. These parts are henceforth referred to as regions. Let us assume that we are interested in some quantity $Q$ for which we would like to determine correlation contributions on top of its HF value. Examples for such a quantity would be the ground state energy $\epsilon$ of a system, the self energy matrix $\Sigma_{ij}(\omega)$, the transmission coefficient $T$ for charge transport through the system under consideration, to name but a few.

An incremental description of the correlation contributions $Q$ to the quantity at hand could start with a correlation calculation in which only excitations inside one of the regions I–VI, e. g. region I in Fig. 2, are allowed. This results in a contribution to the correlation effects which is labeled one–region increment $Q^{\mathrm{I}}$ as expressed in Eq. (10). Of course, there are as many one–region increments as regions chosen to map the system. They are thus indexed with the region they are referring to. In a next step the calculation is repeated with excitations correlating the charge carriers being allowed to a region enlarged by one additional region, for example region II in Fig. 2. The result of this calculation is denoted as $Q^{\mathrm{I,II}}$ and the difference with respect to the one–region increments then isolates the effect of additional excitations involving this additional region II and constitutes the two–region increment as shown in Eq. (11). This procedure can be continued to more and more regions. As an example Eq. (12) shows the three–region increment $Q^{\mathrm{I,II,IV}}$, where the benzene ring and one thiol group are included.

In the end the summation Eq. (13) of all increments is the final approximation to the sought quantity $Q$. The restrictions on the summation indices $A, B$ and $C$ prevent double counting of contributions.

$$\Delta Q^{\mathrm{I}} = Q^{\mathrm{I}} \tag{10}$$

$$\Delta Q^{\mathrm{I,II}} = Q^{\mathrm{I,II}} - \Delta Q^{\mathrm{I}} - \Delta Q^{\mathrm{II}} \tag{11}$$

$$\Delta Q^{\text{I,II,IV}} = Q^{\text{I,II,IV}} - \Delta Q^{\text{I,II}} - \Delta Q^{\text{I,IV}} - \qquad (12)$$
$$\Delta Q^{\text{II,IV}} - \Delta Q^{\text{I}} - \Delta Q^{\text{II}} - \Delta Q^{\text{IV}}$$

$$\boxed{\begin{aligned} Q = \quad & \sum_{A=I}^{IV} \Delta Q^{A} + \\ & \sum_{A>B=I}^{IV} \Delta Q^{A,B} + \\ & \sum_{A>B>C=I}^{IV} \Delta Q^{A,B,C} + \\ & \quad \ldots \end{aligned}} \qquad (13)$$

The main idea of the incremental series (13) is to exploit the mainly local character of correlation corrections to HF results. This feature should manifest itself in a rapid decrease of increments both with the distance between the regions involved and with their number included in the increment. This means that only a few increments need to be calculated. It is crucial to emphasize that the cutoff thus introduced in the summation (13) is well controlled, since the decrease of the incremental series can be explicitly monitored.

Furthermore physical information can be extracted from the incremental scheme. In general the relative weight of specific increments with respect to others helps to identify important correlation contributions as was elaborated upon in earlier works[13,26,27,29].

The special case of the ground state correlation energy $\epsilon$ is well suited to illustrate the incremental scheme. The equations are obtained directly by setting $Q = \epsilon$ in (10)-(12) so that the final approximation reads in analogy to Eq. (13):

$$\epsilon = \sum_{A=I}^{IV} \Delta \epsilon^{A} + \sum_{A>B=I}^{IV} \Delta \epsilon^{A,B} + \sum_{A>B>C=I}^{IV} \Delta \epsilon^{A,B,C} + \ldots \qquad (14)$$

As another application of the incremental scheme, we denote the corresponding equations for the calculation of the self energy. For each matrix element $\Sigma_{ij}(\omega)$ the incremental scheme holds as

$$\Delta\Sigma_{ij}^{A}(\omega) = \Sigma_{ij}^{A}(\omega). \tag{15}$$

$$\Delta\Sigma_{ij}^{A,B}(\omega) = \Sigma_{ij}^{A,B}(\omega) - \Delta\Sigma_{ij}^{A}(\omega) - \Delta\Sigma_{ij}^{B}(\omega). \tag{16}$$

$$\Delta\Sigma_{ij}^{A,B,C}(\omega) = \Sigma_{ij}^{A,B,C}(\omega) - \Delta\Sigma_{ij}^{A,B}(\omega) - \Delta\Sigma_{ij}^{A,C}(\omega) - \Delta\Sigma_{ij}^{B,C}(\omega) - \tag{17}$$

$$\Delta\Sigma_{ij}^{A}(\omega) - \Delta\Sigma_{ij}^{B}(\omega) - \Delta\Sigma_{ij}^{C}(\omega). \tag{18}$$

The self energy is finally approximated by:

$$\Sigma_{ij}(\omega) = \sum_{A}\Delta\Sigma_{ij}^{A}(\omega) + \sum_{A>B}\Delta\Sigma_{ij}^{A,B}(\omega) + \sum_{A>B>C}\Delta\Sigma_{ij}^{A,B,C}(\omega) + \ldots \,. \tag{19}$$

In earlier applications to band structure calculations it was demonstrated that the correlation correction $\gamma = (LUMO - HOMO)_{\mathrm{HF}} - (LUMO - HOMO)_{\mathrm{CORR}}$ of the band gap gives a suitable measure of the correlation effects accounted for, where $(LUMO - HOMO)_{\mathrm{HF}}$ is the HOMO-LUMO gap on the HF level and $(LUMO - HOMO)_{\mathrm{CORR}}$ is the correlated result. This correction can also be used as a target quantity for the incremental scheme in very much the same spirit as the ground state correlation energy $\epsilon$ inserted above.

Plenty of experience has been gained with applications of the incremental scheme to ground state properties and very good convergence of the incremental scheme in all cases was found[9–11,13,14].

In an earlier work we pointed out that Epstein–Nesbet second order perturbation theory (EN2) results are obtained if the diagonalization of the Hamiltonian matrix (5) is skipped and its diagonal elements used as eigenvalues in Eq. (7) instead[25]. Those eigenvalues are then given by:

$$\lambda_{\mathrm{rsa}}^{\mathrm{R}} = F_{\mathrm{rr'}} + F_{\mathrm{ss'}} - F_{\mathrm{aa'}} + \Gamma(\mathrm{rs;rs}) - \Gamma(\mathrm{ra;ra}) - \Gamma(\mathrm{sa;sa}) \tag{20}$$

$$= \epsilon_{\mathrm{r}} + \epsilon_{\mathrm{s}} - \epsilon_{\mathrm{a}} + J_{\mathrm{rs}} - K_{\mathrm{rs}} - J_{\mathrm{ra}} + K_{\mathrm{ra}} - J_{\mathrm{sa}} + K_{\mathrm{sa}}.$$

In a further simplification the ordinary Møller–Plesset perturbation theory result (PT2) is obtained by only retaining the Fock quantities: $\lambda_{\mathrm{rsa}}^{\mathrm{R}} = \epsilon_{\mathrm{r}} + \epsilon_{\mathrm{s}} - \epsilon_{\mathrm{a}}$. A comparison of algebraic expressions and diagrams then allowed to establish the relation

$$\Sigma_{ij}^{(\mathrm{PT2})}(\omega = \epsilon_{\mathrm{i}}) = H_{ij}^{\mathrm{eff},(\mathrm{PT2})} - F_{ij}, \tag{21}$$

where $H_{ij}^{\text{eff},(\text{PT2})}$ is the second order effective Hamiltonian. In a similar fashion we can show that the advanced part of the self energy in Eq. (6) to second order perturbation theory taken at the frequency $\omega = \epsilon_i$ can be used to find the second order value of the ground state correlation energy. To this end we evaluate this part of the self energy for indices $n, m$ referring to virtual HF orbitals $r, s$. Then the second order expression for $\Sigma_{nm}^A$ reads in analogy to Eq. (7)

$$\Sigma_{rs}^A(\omega = \epsilon_r) \approx \sum_{abt} \Gamma(ab; rt) \frac{1}{(\epsilon_r + \epsilon_t - \epsilon_a - \epsilon_b)} \Gamma(ab; st). \tag{22}$$

Taking the trace of (22) yields

$$\sum_{abtr} \Gamma(ab; rt) \frac{1}{(\epsilon_r + \epsilon_t - \epsilon_a - \epsilon_b)} \Gamma(ab; rt). \tag{23}$$

This is just the formula for the second order diagram for the ground state correlation energy as for example given by Monkhorst[31] or Lindgren and Morrison[32]. In consequence at the perturbative level the correlation contribution $\epsilon$ to the ground state energy can easily be extracted from the self energy formulars and does not need to be calculated via the cumbersome standard frequency integral over the Green's function ($E_0 = -i \lim_{\eta \to 0^+} \int_{-\infty}^{+\infty} \frac{d\omega}{2\pi} e^{i\eta\omega} \sum_m \left[ \left( F + \frac{1}{2}\Sigma(\omega) \right) G(\omega) \right]_{mm}$ [24]).

Despite earlier applications of the incremental scheme to the self energy and hence to the correlation correction of excitation energies[24–26], the scheme is rendered awkward due to the frequency dependence as is illustrated in the next section.

### C. A frequency independent incremental scheme

We start from the observation that the frequency dependence introduced by the frequency dependent energy denominator in Eq. (7) is responsible for the incremental scheme to become cumbersome and infeasible for two reasons. First of all the sum in Eq. (7), (9) runs over the full matrix of the correlation space (in the explicit formula the correlation space is described by indices $r, s, a$, or in the basis diagonalizing the Hamiltonian is abbreviated as $q$) which can easily extend to some million single terms. This sum has to be evaluated for each frequency $\omega$ separatley for any $\omega$ the process comes accross in the iterations. Secondly this has to be done not just for one self energy matrix, but rather for all matrices involved in the incremental description of the self energy. These summations are thus very time consuming. Moreover, to minimize this time, all the eigenvectors $D_{n,q}^R$ and eigenvectors $\lambda_q^R$ in Eq. (9) should be simultaneously available in the main memory of the computer for all increments, which means that the main memory requested is also very huge.

An accurate variable separation of the denominator would surly serve as a remedy to this shortcoming. However, weight has to be put on the condition that the separation be mathematically exact and numerically manageable.

First we cast the the problem into an abstract form. The culprit is of the form

$$\frac{1}{\omega - \lambda}, \tag{24}$$

where $\omega$ is the frequency for which the self energy is to be evaluated, and $\lambda$ is some eigenvalue of higher excitations. In the case of the 2p1h space, we have the relation

$$0 < \omega < \lambda, \tag{25}$$

while in the case of the 2h1p space this becomes

$$\lambda < \omega < 0. \tag{26}$$

We are looking for a way to decompose this expression into different factors for $\omega$ and $\lambda$, ideally in the form

$$\frac{1}{\omega - \lambda} = \tilde{f}\tilde{g}(\lambda)\tilde{h}(\omega), \tag{27}$$

were $\tilde{f}$ is a constant and $\tilde{g}$ and $\tilde{h}$ are some suitable functions. This decomposition would allow to first evaluate the costly summations over the correlation space once and for all independent of the frequency and in the end simply multiply with $\tilde{h}(\omega)$ to obtain the full expression at the desired frequency.

In the following we demonstrate how an ingenious decomposition developed by Hackbush and Khoromskij[33] could be used for the problem at hand. They have looked at an expression similar to (24) and found the mathematically exact formula

$$\frac{1}{x+y} = \int_0^\infty d\rho f(\rho)\tilde{g}(x,\rho)\tilde{g}(y,\rho). \tag{28}$$

Clearly this would solve the problem in the sense that we could take the integral outside the summation in the self energy and first evaluate all the $f(\rho), g(y,\rho)$ for all $\rho \in [0,\infty]$ with the idea $y = -\lambda$ and then evaluate $g(x,\rho)$ in the same interval for $\rho$ with $y = \omega$. However, the decisive point to note is that the evaluation of the integral should be much easier than the summation over the frequency dependent self energy expression, otherwise we would have gained nothing. In this respect a simple minded version of Eq. (28) would turn out to fail. The situation is saved by the discovery of Hackbush and Khoromskij who established that the following version is absolutely reliable and rapidly convergent in all cases:

$$f(\rho) = \frac{cosh(\rho)}{1 + e^{-sinh(\rho)}} \tag{29}$$

$$\tilde{g}(x, \rho) = e^{-xg(\rho)} \tag{30}$$

$$g(\rho) := log(1 + e^{sinh(\rho)}). \tag{31}$$

As is obvious from Eq. (31), we must have $x, y > 0$. In fact Hackbush and Khoromskij formulate as condition for convergence:

$$1 < x < y < \infty. \tag{32}$$

We will show later, how the problem at hand can be recast to be in conformance with Eq. (32). Here we first point out the tremendous simplification the decomposition can bring about. The somewhat involved form of the integrand (29) finds its justification in the simple and stable form, in which the integrand can be evaluated. In fact we have:

$$\int_0^\infty d\rho f(\rho) e^{-xg(\rho)} e^{-yg(\rho)} \approx h \sum_{m=-l}^{l} f(hm) e^{-xg(hm)} e^{-yg(hm)}, \tag{33}$$

where $l$ defines the level of approximation and the step width $h$ is given by

$$h = \frac{1}{l} log\left(\frac{4\pi^2 l}{3}\right). \tag{34}$$

(In fact we have used the formula from ref.[34] with $b = 1/2$, $a = 1$ and $\delta = \pi/3$). We will establish in the next section that a value as small as $l = 64$ is sufficient to achieve machine precision in the approximation (33), while $l = 32$ is perfectly satisfactory for the final result as well.

Before rewriting the incremental scheme using the decomposition (33), we first reformulate the original problem so as to match condition (32). To do so four cases have to be distinguished:

*1. The retarded part for positive frequencies.*

When the retarded part of the self energy is evaluated for positive frequencies, then $\omega$ normally runs over the range of low lying virtual HF orbitals. This is the regime, where the application of the incremental scheme is admissible and where we are interested in obtaining correlation corrections. The eigenvalues obtained from diagonalizing the Hamiltonian $H^R$, on the other hand, describe single and double excitations on top of $N+1$–particle HF determinants, and have thus higher energies than the low–lying $N+1$–particle HF states themselves. The situation is pictured in Fig. 1 a).

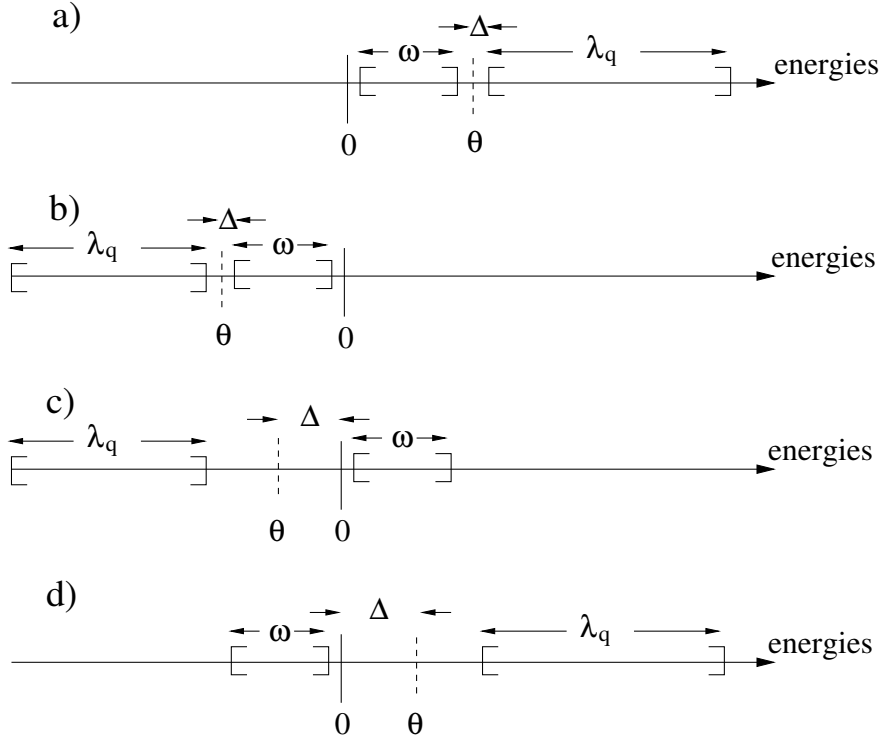

FIG. 1: *Illustration of the energy ranges in the four different cases in the setup of the self energy as explained in the text.*

The index $q$ just labels the eigenvalues in the correlation space as in Eq. (9). With the restriction $\omega < \omega^{\max} < \lambda_q^{\min} < \lambda_q$ and the definitions

$$\theta := \frac{\lambda_q^{\min} + \omega^{\max}}{2}, \tag{35}$$

$$\Delta := \lambda_q^{\min} - \theta \tag{36}$$

we can rewrite the energy denominators as:

$$\frac{1}{\omega - \lambda_q} = -\frac{1}{\Delta} \frac{1}{\frac{\lambda_q - \theta}{\Delta} + \frac{\theta - \omega}{\Delta}} \tag{37}$$

$$= -\frac{1}{\Delta} \frac{1}{x + y}. \tag{38}$$

In the last line correspondence to the mathematical treatment above is made by the identifications

$$x := \frac{\theta - \omega}{\Delta} \quad (39)$$

$$y := \frac{\lambda_q - \theta}{\Delta} \quad (40)$$

together with the observation that we always fulfill

$$1 < x < y < \infty. \quad (41)$$

With this correspondence established, we can simply state the results of rewriting the energy denominator in the remaining three cases.

2. *The advanced part for negative frequencies.*

According to panel b) of Fig. 1 frequencies and eigenvalues are restricted as $\lambda_q < \lambda_q^{\max} < \omega_{\min} < 0$ and the definitions

$$\theta := \frac{\lambda_q^{\max} + \omega^{\min}}{2}, \quad (42)$$

$$\Delta := \theta - \lambda_q^{\max} \quad (43)$$

lead to

$$\frac{1}{\omega - \lambda_q} = \frac{1}{\Delta} \frac{1}{\frac{\omega - \theta}{\Delta} + \frac{\theta - \lambda_q}{\Delta}} \quad (44)$$

$$= \frac{1}{\Delta} \frac{1}{x + y}, \quad (45)$$

where again the identification

$$x := \frac{\omega - \theta}{\Delta} \quad (46)$$

$$y := \frac{\theta - \lambda_q}{\Delta} \quad (47)$$

guarantees condition (41).

3. *The advanced part for positive frequencies.*

According to panel c) of Fig. 1 frequencies and eigenvalues are now restricted by $\lambda_q < \lambda_q^{\max} < 0 < \omega_{\min} < \omega$ and the definitions

$$\theta = \frac{\lambda_q^{\max}}{2} \ , \ \Delta = \frac{|\theta|}{2} \quad (48)$$

lead to

$$\frac{1}{\omega - \lambda_q} = \frac{1}{\Delta} \frac{1}{\frac{\omega+|\theta|}{\Delta} + \frac{|\lambda_q|-|\theta|}{\Delta}} \tag{49}$$

$$= \frac{1}{\Delta} \frac{1}{x+y}, \tag{50}$$

while again the identification

$$x := \frac{\omega + |\theta|}{\Delta} \ , \ y := \frac{|\lambda_q| - |\theta|}{\Delta} \tag{51}$$

fulfills condition (41).

*4. The retarded part for negative frequencies.*

Finally panel d) of Fig. 1 illustrates that frequencies and eigenvalues are now restricted by $< \omega < \omega_{\min} < 0 < \lambda_q^{\min} < \lambda_q$ and the definitions for $\theta$ and $\Delta$ now read

$$\theta = \frac{\lambda_q^{\min}}{2} \ , \ \Delta = \frac{\theta}{2} \tag{52}$$

so that

$$\frac{1}{\omega - \lambda_q} = -\frac{1}{\Delta} \frac{1}{\frac{|\omega|+\theta}{\Delta} + \frac{\lambda_q - \theta}{\Delta}} \tag{53}$$

$$= -\frac{1}{\Delta} \frac{1}{x+y}, \tag{54}$$

where $x, y$ correspond to:

$$x := \frac{|\omega| + \theta}{\Delta} \ , \ y := \frac{\lambda_q - \theta}{\Delta}. \tag{55}$$

This identification again fulfills condition (41).

**D. The frequency independent formulas**

We now proceed with rewriting the formulas for the incremental scheme for the case of the retarded part of the self energy while looking at positive frequencies. The other parts can be reformulated in strictly analogous ways. A single energy denominator is now evaluated as:

$$\frac{1}{\omega - \lambda_q} = -\frac{1}{\Delta} h \sum_{m=-l}^{l} f(hm) e^{-\frac{\theta-\omega}{\Delta} g(hm)} e^{-\frac{\lambda_q - \theta}{\Delta} g(hm)}, \tag{56}$$

where the summation over $l$ is a very finite one with not more than about 100 parts, independent of the quantum chemical size of the problem. This decomposition inserted in the expression for one self energy matrix element in Eq. (7) then takes the form (suppressing the mathematical convergence factor $i\delta$):

$$\begin{aligned}
\Sigma^{R}_{nm}(\omega) &= \sum_{rsa;r's'a'} \Gamma(\text{rs; na}) \sum_{q} S^{R}_{rsa;q} \frac{1}{\left(\omega - \lambda^{R}_{q} + i\delta\right)} S^{R}_{q;r's'a'} \Gamma(\text{r's'; ma'}) \\
&\approx \sum_{rsa;r's'a'} \Gamma(\text{rs; na}) \sum_{q} S^{R}_{rsa;q} \left(\frac{-h}{\Delta}\right) \sum_{m=-l}^{l} f(hm) e^{-\frac{\theta-\omega}{\Delta} g(hm)} e^{-\frac{\lambda^{R}_{q}-\theta}{\Delta} g(hm)} S^{R}_{q;r's'a'} \Gamma(\text{r's'; ma'}) \\
&= \sum_{m=-l}^{l} e^{-\frac{\theta-\omega}{\Delta} g(hm)} \left[ \sum_{rsa;r's'a'} \Gamma(\text{rs; na}) \sum_{q} S^{R}_{rsa;q} \left(\frac{-h}{\Delta}\right) f(hm) e^{-\frac{\lambda^{R}_{q}-\theta}{\Delta} g(hm)} S^{R}_{q;r's'a'} \Gamma(\text{r's'; ma'}) \right] \\
&= \sum_{m=-l}^{l} e^{-\frac{\theta-\omega}{\Delta} g(hm)} \left[ \sum_{q} D^{R}_{n;q} \left(\frac{-h}{\Delta}\right) f(hm) e^{-\frac{\lambda^{R}_{q}-\theta}{\Delta} g(hm)} D^{R}_{q;m} \right],
\end{aligned} \qquad (57)$$

where the summation describing the decomposition of the energy denominators has been taken to be the outermost summation and again the stored quantities (8) have been used. The quantities in the angled brackets are frequency independent. For each value of $l$ we can thus calculate independent of the frequency beforehand and then store a matrix $\Theta^{R,l}_{nm}$ given by

$$\Theta^{R,l}_{nm} = \sum_{q} D^{R}_{n;q} \left(\frac{-h}{\Delta}\right) f(hm) e^{-\frac{\lambda^{R}_{q}-\theta}{\Delta} g(hm)} D^{R}_{q;m} \qquad (58)$$

The incremental scheme then carries over in full to this new quantity and reads:

$$\Delta \Theta^{A}_{ij} = \Theta^{A}_{ij}. \qquad (59)$$

$$\Delta \Theta^{A,B}_{ij} = \Theta^{A,B}_{ij} - \Delta \Theta^{A}_{ij} - \Delta \Theta^{B}_{ij}. \qquad (60)$$

$$\Delta \Theta^{A,B,C}_{ij} = \Theta^{A,B,C}_{ij} - \Delta \Theta^{A,B}_{ij} - \Delta \Theta^{A,C}_{ij} - \Delta \Theta^{B,C}_{ij} - \qquad (61)$$
$$\Delta \Theta^{A}_{ij} - \Delta \Theta^{B}_{ij} - \Delta \Theta^{C}_{ij}. \qquad (62)$$

$$\Theta_{ij} = \sum_{A} \Delta \Theta^{A}_{ij} + \sum_{A>B} \Delta \Theta^{A,B}_{ij} + \sum_{A>B>C} \Delta \Theta^{A,B,C}_{ij} + \ldots \quad . \qquad (63)$$

Whenever the frequency dependent self energy needs to be evaluated, it can then easily be assembled as

$$\Sigma^{R}_{nm}(\omega) = \sum_{m=-l}^{l} e^{-\frac{\theta-\omega}{\Delta} g(hm)} \Theta^{R,l}_{nm}. \qquad (64)$$

This can be done in no time, since only very few $l$–terms are needed, and constitutes the breakthrough towards a feasible incremental scheme of the self energy. The incremental scheme is now applied to $\Theta_{nm}^{R,l}$ irrespective of the frequency, and the self energy can then be obtained in the end by virtue of (64).

This frequency independent formalism has now been fully programmed as an add–on option in the program package GREENS and allows to test the numerical performance as put forth in the following section.

## III. RESULTS AND DISCUSSION

In this work we want to demonstrate the numerical feasibility and stability of the decomposition of energy denominators when applied to an incremental description of the self energy. To this end we compare the results obtained according to formula (64) with an earlier frequency dependent procedure described by Eq. (7). All self energies are evaluated in the EN2 scheme. We focus on a system which we recently investigated in the frame of electronic transport through molecular junctions[27]. The molecule under consideration is a para–ditholbenzene. A vdz basis set with polarization functions was used throughout the calculations. In a first preparatory step localized HF orbitals were obtained for the molecules employing the Pipek–Mezey option of the program package MOLPRO[35]. The four–index transformation was accomplished by means of the HF program package WANNIER[36] and the subsequent correlation calculations were performed by the program GREENS developed in our laboratory[24,25,27].

The incremental scheme has been applied to the para–ditholbenzene molecule with a partitioning shown in Fig. 2. The thiol groups are assembled into incremental regions I and III. For illustrative purposes the carbon ring was split into two asymmetric parts, a larger one (region IV, upper part) and a smaller one in the lower part, denoted as part II. The incremental scheme will then display how the different parts contribute to the correlation effects.

### A. Numercial Reliability

We start with a numerical analysis of the validity of the proposed energy denominator decomposition (56). To this end we have chosen the spectrum of the 2p1h space of a third of the benzene ring (one–region increment comprising of region II in Fig. 2). For each of the approximately 50000 eigenvalues $\lambda_q$ the expression $\frac{1}{\omega - \lambda_q}$ was evaluated for $w = 0.114048\ Hartree$. $\lambda_q$ was found to be in the interval $0.383114 \leq \lambda_q \leq 7.347588\ Hartree$.

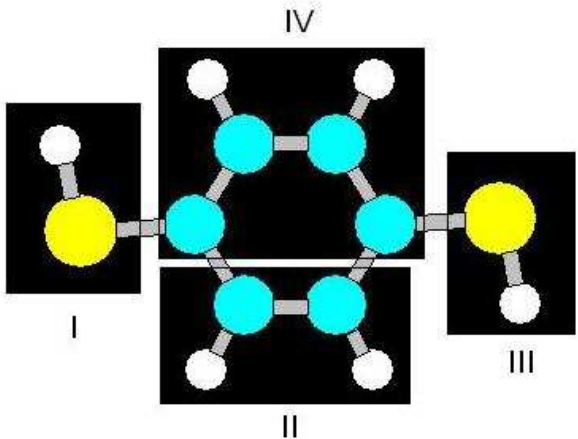

FIG. 2: *Sketch of the incremental scheme, exemplifying a possible partition of the para–dithiolbenzene comprising four parts denoted as I, II, III and IV.*

The relative error of the approximated denominator with respect to the numerically true result is plotted for all eigenvalues $\lambda$ for various values of $l$ in Fig. 3.

From the upper left panel of the figure one can conclude that a value of $l = 8$ leads to relative errors of up to 4 %. However, this error is down to 0.04 % for $l = 16$ as shown in the upper right panel. The choice $l = 32$ brings this down to a relative error of $10^{-8}$ (lower left panel) and finally machine precision $10^{-15}$ is achieved for $l = 64$ (lower right panel).

The relative error can be summed up for all the eigenvalues for this increment, which are some ten thousands, so the machine error of $10^{-15}$ should then amount to some $10^{-11}$. As can be seen from Tab. I, this is indeed the case.

TABLE I: *Integrated relative error of the denominator decomposition for various levels of accuracy $l$.*

| $l =$ | 8 | 16 | 32 | 64 |
|---|---|---|---|---|
| integrated relative error | 352.6939 | $-0.7252166$ | $1.401267 * 10^{-6}$ | $-2.706844 * 10^{-12}$ |

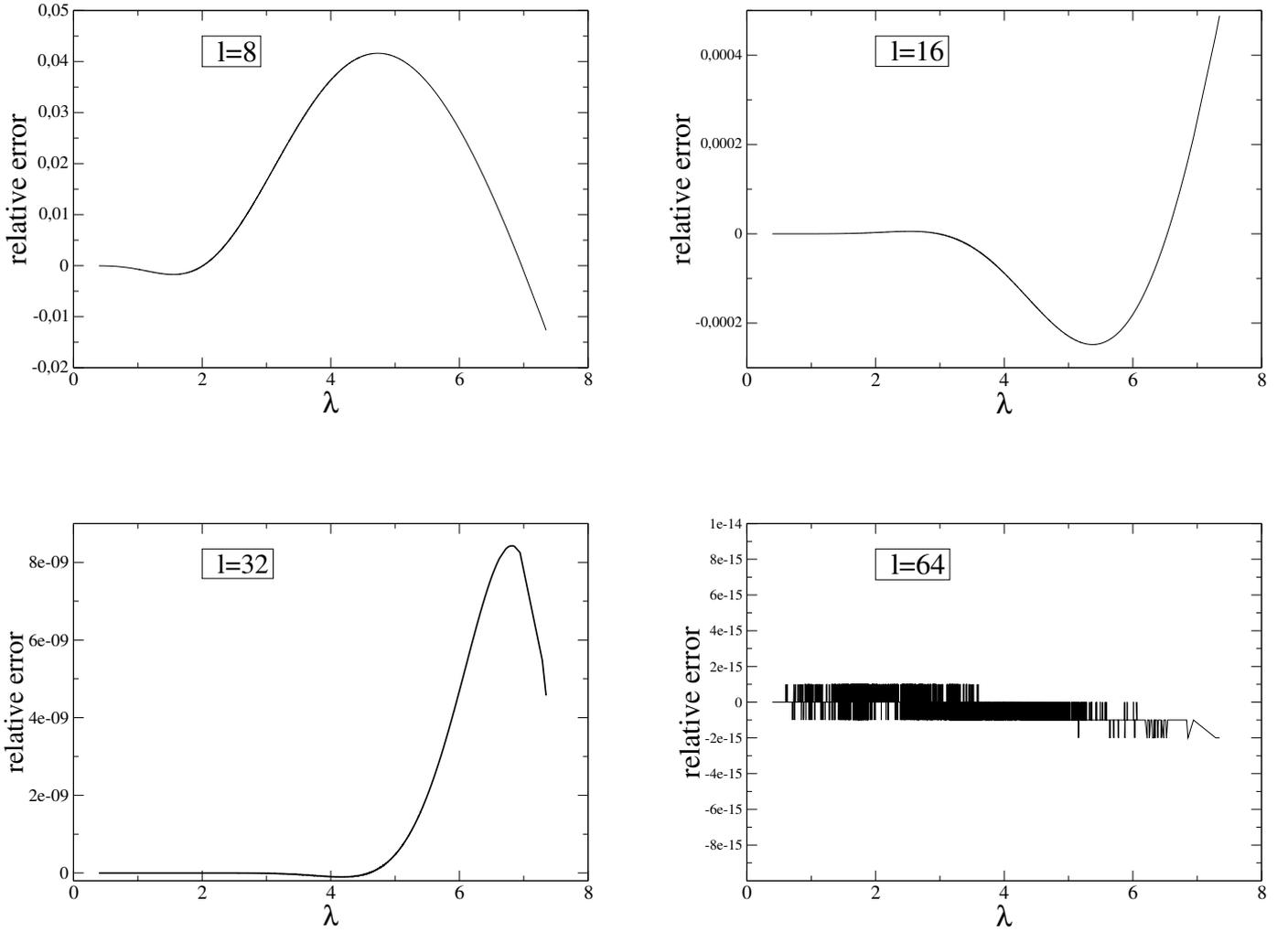

FIG. 3: *Plots of the relative error of the denominator decomposition for all eigenvalues $\lambda$. The eigenvalues are given in Hartree. The range $l$ of the approximate sum of the decomposition is $l = 8$ (upper left panel), $l = 16$ (upper right panel), $l = 32$ (lower left panel), $l = 64$ (lower right panel),*

For $l = 64$ the integrated relative error amounts to $-2.7 * 10^{-12}$. This quantity is a good measure for the error of the method, because the calculation of the self energy precisely sums over the same amount of denominators. Thus all data are intrinsically only accurate to $10^{-11}$, irrespective of whether or not our proposed decomposition is applied. For $l = 32$ all energies would come out accurate to $10^{-6}$, which is still acceptable, while lower values of $l$ should not be used. In sum the performance of the denominator decomposition is very impressive. $l = 64$ implies that the energy denominators are approximated with 129 terms only to achieve machine precision over the entire range of possible values that might occur.

Finally the convergence with $l$ is checked for the one–region increment II and for the region including parts II and IV. The latter comprises of the carbon ring system. The results are displayed in Tab. II. The gap correction $\gamma$ denotes the correlation corrections to the HF HOMO–LUMO gap in eV. This quantity is frequently used in discussing the incremental scheme so as to assess correlation effects in one single number rather than having to monitor huge matrices[24–27]. (This is not to say that we propagate a particular physical meaning for the HOMO-LUMO gap of a finite system). In addition we have calculated the incremental contributions to the ground state energy. Two increments were considered, the one–region increment $I^{II}$ of region $II$ in Fig. 2 and the sum of increments for the two regions $II$ and $IV$ together, $I^{II} + I^{IV} + \Delta I^{II,IV}$, which describes the full benzene ring. Comparing the numerical values obtained for $l = 8, 16, 32, 64$ and the exact result gives a clear idea about the convergence of the procedure with respect to $l$.

TABLE II: *Convergence of increments with respect to $l$. Both the gap correction $\gamma$ (in eV) and the correlation contribution to the ground state energy $\epsilon$ (in Hartree) are shown for the increment referring to region II and the increment referring to regions II and IV.*

|  | gap correction $\gamma$ | | ground state correction $\epsilon$ | |
| --- | --- | --- | --- | --- |
|  | $I^{II}$ | $I^{(II,IV)}$ | $I^{II}$ | $I^{(II,IV)}$ |
| l=8 | 1.21641625693881 | 3.21946344784797 | -0.184182392681587 | -0.624656406117148 |
| l=16 | 1.23811508294961 | 3.27386787876785 | -0.184205522065484 | -0.625769114870404 |
| l=32 | 1.23809733469899 | 3.27381320151962 | -0.184205739664410 | -0.625769554246712 |
| l=64 | 1.23809733357899 | 3.27381319929349 | -0.184205739664344 | -0.625769554245726 |
| exact | 1.23809733357899 | 3.27381319929352 | | |

Convergence is achieved for $l = 32$, where the gap correction already agrees with the exact result up to $10^{-7}$ $eV$ for both increments. Machine precision ($10^{-11}$ $eV$) is obtained for $l = 64$. This tendency is found both for the one–region increment $I^{II}$ (second column in Tab. II) and for the sum of regions II and IV, $I^{II} + I^{IV} + \Delta I^{II,IV}$ (third column). Comparing the cases $l = 32$ and $l = 64$ demonstrates that the convergence for the ground state quantity $\epsilon$ is even more convergent, though it should be noted that numbers are in Hartree in this case rather than eV: The change brought about by increasing $l$ from 32 to 64 yields a change in the order of $10^{-13}$ and $10^{-12}$ $Hartree$ for the ground state case and the two increments, respectively, while these changes are $10^{-13}$ and $10^{-9}$ $eV$ for the gap correction. The agreement between the $l = 64$ decomposition and the exact result (obtained from using the frequency dependent denominators) is very impressive. The same results are obtained for the first increment while the deviation for the second one is $3 * 10^{-14}$ $eV$. Thus the real relative errors are much smaller than the worst possible one, which was calculated above to be in the order of $10^{-11}$. This can be understood by looking again at the lower right panel in Fig. 1 which gives the relative error of a single energy denominator decomposition for $l = 64$. The error is quantized with the machine precision $10^{-15}$, but the sign is basically equally distributed between $+$ and $-$, so that the sum over the real relative errors is basically avaraged out to zero. This can be understood on statistical grounds. By contrast, in Tab. I we discussed the worst case, a sum of the absolute value of the individual errors.

In what follows the incremental scheme is analyzed with the parameter $l = 64$ throughout. The results are shown in Tab. III for both the correlation correction $\gamma$ to the HF HOMO–LUMO gap and the ground state energy. The latter was obtained in the EN2 approximation from a formula similar to (23). However, it was obtained in the same manner as the self energy after the decomposition of the denominator was made. The terms 'one–region', 'two–region' and 'three–region' refer to inclusion of one–region, two–region and three–region increments. The abbreviation $nn$ indicates next–nearest neighbour constellations. From the table it can be seen that most correlation corrections are assessed with the one–region increments, which correct the HF value of the HOMO-LUMO gap of $11.309$ $eV$ by an amount of $3.845$ $eV$. Inclusion of two–region increments leads to an additional correction of $0.779$ $eV$. As was found in an earlier work, three–region increments do not significantly contribute at all. The next–nearest–neighbour three–region contributions including the full benzene ring only amount to $0.042$ $eV$. Even the second–next–nearest neighbour two–region increment between regions I and III, i. e. the two thiol groups, amounts to a meagre $0.0243 = 0.7787 - 0.7544$ $eV$, as can be seen by subtracting the two incremental values for 'nn two–region' and 'all two–region' in column '$\Delta_\gamma$' in Tab. III. The overall convergence of the incremental scheme becomes apparent by looking at the numbers printed bold face. They state the contributions for up to all one– all two– and the next–nearest neighbour three–region increments which include the full benzene ring.

The influence of the next–nearest–neighbour increments (regions I+II, I+IV and II+IV) are stated separately in the lines marked 'I+II', 'I+IV' and 'II+IV' in the table. The column '$\Delta_\gamma$' indicates the improvement obtained upon including the respective two–region increment *in addition to* all one–region increments. It is rather small for the 'I+II' case while the largest two–region increment is the 'II+IV' increment which restores the integrity of the benzene ring for the correlation treatment.

As can be seen from the last column in Tab. III, these findings are confirmed for the correlation contribution to the ground state energy in very much the same way. The convergence is even faster than for the gap correction. The one–region increments yield a correlation energy of $-0.7725$ $Hartree$, the two–region increments amount to an additional contribution of $-0.2096$ $Hartree$ and the three–region increments do not contribute at all. The only two–region increment which is not nearest neighbour, i. e. the two–region increment I,III also yields a negligible contribution with just $0.0007$ $Hartree$.

The overall result is a correlation contribution of $-0.982$ $Hartree$ in EN2. This result is embedded by the results of ordinary Møller–Plesset second order perturbation theory (PT2), which were obtained using MOLPRO. PT2 using the very same localized orbitals yields a correlation contribution $\epsilon$ to the ground state energy of $\epsilon = -0.618$ $Hartree$, using the canonical HF orbitals increases this result to $-1.016$ $Hartree$. As expected we thus recover the fact that PT2 in local orbitals yields results which fall way short of those where canonical orbitals are used. However, the EN approximation in local orbitals allows to make up for most of this shortcoming. This has been analyzed time and again by various authors[20,37,38].

In sum it can be concluded that converged results are already obtained on the two–region increment level and would justify the cutoff applied to the incremental series after the two–region contributions. In this way the local character of a correlation hole around a quasi particle manifests itself[39]. It is important to note that the incremental scheme thus allows for a well–monitored and significant simplification in the numerical effort, since the space to describe two–region increments is in general significantly smaller than the one for three regions included.

### B. Computational Savings

We now turn to a discussion of the computational savings achieved by the new formulation of the incremental scheme for the self energy. In particular we compare computation time and memory requirements for both the case of the common frequency dependent treatment (denoted $\Sigma(\omega)$) and for the case that the indermediate frequency independent quantity $\Theta$ is evaluated and then the self energy restored (denoted $\Theta$). In the latter case again all calculations are done with machine precision, meaning $l = 64$. Since both computation time and memory requirement scale linearly with $l$, the computational effort for other choices of $l$ can be obtained accordingly. Tab. IV compares computation times and memory requirements for the main memory RAM. This is done for an incremental scheme comprising of a) just the one–region increments, b) additionally also the next–nearest neighbour two–region increments and c) all one– and two–region increments plus the two next–nearest neighbour three–region increments which contain the full benzene ring. There are 4, 9 and 12 increments for the cases a), b) and c), respectively.

TABLE III: *Incremental scheme for the ground state correlation energy ϵ in Hartree and for the correlation correction γ to the HOMO-LUMO gap (in eV) calculated at a convergence constant $l = 64$. Row 'all one–region' gives the correlated results with all one–region increments included, row 'nn two–region' also takes into account all nearest neighbour two–region increments and so on. $\Delta_\epsilon$ and $\Delta_\gamma$ monitor the incremental improvement.*

| increment | narrowing of the gap | $\Delta_\gamma$ | $\epsilon$ | $\Delta_\epsilon$ |
|:---:|:---:|:---:|:---:|:---:|
| HF | gap=11.3088 | | | |
| all one–region | **3.8452** | **3.8452** | **−0.7725** | **−0.7725** |
| I+II | 3.9162 | 0.071 | -0.7939 | -0.0214 |
| I+IV | 4.0754 | 0.2302 | -0.8043 | -0.0317 |
| II+IV | 4.1561 | 0.3109 | -0.8256 | -0.0531 |
| nn two–region | 4.5996 | 0.7544 | -0.9815 | -0.2089 |
| all two–region | **4.6239** | **0.7787** | **−0.9821** | **−0.2096** |
| nn three–region | **4.6659** | **0.0420** | **−0.9821** | **−0.0000** |

The column 'Speed–up' denotes how many times faster the frequency independent treatment $\Theta$ performs compared to the conventional frequency dependent procedure. The last column states the relative savings of main memory RAM in percent. In the $\Theta$ case the computation time is almost perfectly linear with the number of increments, increasing from about 70 seconds for 4 increments to almost 210 seconds for 12 increments. This is natural since while running the incremental scheme the incremental self energy matrix is assembled from $(2l + 1)$ $\Theta$ matrices according to Eq. (64), irrespective of the nature of the increments involved. This is contrasted by the behaviour of the computation time in the $\Sigma(\omega)$ case, where for each increment as many terms have to be assembled as the correlation space covered by this increment might contain as is visible from Eq. (8). As the number of increments considered on a certain accuracy level gets larger, so does the number of regions contained in the additional increments. But this entails a tremendous increas in the size of the correlation space. The doubling of computation time from case b) with 9 increments to case c) with just three increments more illustrates this point. In the present case a speed–up of a factor 47 was found. The savings on main memory RAM are also significant. In the $\Theta$ case the memory space needed increases linearly with the number of increments for the same reason as does the computation time. The quantities to be stored intermediately are just the $\Theta$ matrices for each increment. On the other hand in the $\Sigma(\omega)$ case the eigenvectors $D_{n,q}^{R}$ and eigenvalues $\lambda_q^R$ from Eq. (8) have to be stored for each individual increment, where $q$ again runs over the entire correlation space of the specific increment.

TABLE IV: *Numerical effort for the frequency dependent treatment ($\Sigma(\omega)$) and the frequency independent calculations ($\Theta$) in the three cases a), b) and c) as specified in the text. Both computation time (in s) and main memory RAM requirement (in MB) are specified. The second column gives the number of increments involved in the calculation. Column 'Speed–up' states the number of times the $\Theta$ case is faster and column 'Savings' states the amount of RAM in % saved by the $\Theta$ case compared to the $\Sigma(\omega)$ case.*

| case | no. of increments | time/s | | | RAM/MB | | |
|---|---|---|---|---|---|---|---|
| | | $\Sigma(\omega)$ | $\Theta$ | Speed–up | $\Sigma(\omega)$ | $\Theta$ | Savings |
| a) | 4 | 1509 | 69 | 22 | 130 | 106 | 18% |
| b) | 9 | 5199 | 134 | 39 | 477 | 240 | 50% |
| c) | 12 | 9819 | 207 | 47 | 817 | 322 | 61% |

In sum the strictly linear scaling of time and memory requirements in the $\Theta$ case leads to a speed–up of about a factor 50 and memory savings of almost two thirds. Also because of the linear scaling the speed–up and memory savings can be expected to be even much larger for larger systems, as is indicated by the tendencies displayed by Tab. IV.

## IV. CONCLUSIONS

This work introduces a new, frequency independent incremental scheme based on local HF orbitals to construct the self energy and ultimately the Green's function with correlations included in an *ab initio* way. This is achieved by first applying the incremental scheme to a frequency independent matrix, thus recovering the full numerical savings usually provided by the incremental scheme. Subsequently the frequency dependent self energy and Green's function can be constructed without loss of numerical precision, yet without additional numerical costs. This is brought about by a suitable decomposition of the energy denominators in a frequency dependent and an independent part. In a detailed analysis we demonstrated that no numerical errors occur during this procedure. The application of the incremental scheme is speeded up by a factor of 50.

In sum the presented approach allows to efficiently assess correlation effects in general heterogenous systems by means of the standard Green's function theory in combination with the incremental scheme.

We believe that this is a breakthrough which will lead to large scale applications in the future.

## V. ACKNOWLEDGEMENTS

The author is grateful for an inspiring discussion with Boris Khoromskij from the Max–Planck–Institute for Mathematics in the Sciences (Leipzig/Germany). Support from the German Research Foundation (Deutsche Forschungsgemeinschaft DFG) in the frame of the priority program SPP 1145 is also appreciated.

---